\documentclass[12pt,a4paper]{article}

\usepackage[utf8]{inputenc}
\usepackage[T1]{fontenc}
\usepackage{amsmath,amssymb,amsthm}
\usepackage{graphicx}
\usepackage{hyperref}
\usepackage{cite}
\usepackage{geometry}
\usepackage{float}
\usepackage{booktabs}
\usepackage{caption}
\usepackage{subcaption}
\usepackage{physics}
\usepackage{slashed}
\usepackage{bbold}
\usepackage{dsfont}
\usepackage{mathrsfs}
\usepackage{empheq}
\usepackage{xcolor}
\usepackage{enumitem}
\usepackage{titlesec}
\usepackage{fancyhdr}
\usepackage{abstract}
\usepackage{multirow}
\usepackage{longtable}
\usepackage{appendix}

\theoremstyle{definition}

\theoremstyle{remark}

\newcommand{\lag}{\mathcal{L}}

\newcommand{\act}{\mathcal{S}}
\newcommand{\EM}{\text{EM}}

\newcommand{\PP}{\mathcal{P}}

\newcommand{\HH}{\mathcal{H}}
\newcommand{\GG}{\mathcal{G}}
\newcommand{\FF}{\mathcal{F}}

\newcommand{\Ocal}{\ensuremath{\mathcal{O}}}

\title{\textbf{Black Hole Solutions with Electric and Magnetic Charges in Nonlinear Electrodynamics}}
\author{
    \textbf{Shu Luo}\\
    \textit{School of Physics, University of Science and Technology of China}\\
    \textit{Email: ls040629@mail.ustc.edu.cn}
}
\date{\today}

\begin{document}

\maketitle

\begin{abstract}
This review article provides a comprehensive and self-contained overview of black hole solutions coupled to nonlinear electrodynamics (NLED) with both electric and magnetic charges. We systematically discuss the theoretical foundations, including the general action principle, the Hamiltonian $P$-framework for constructing exact solutions, and the classification of NLED theories (Born--Infeld, Euler--Heisenberg, power-law, logarithmic, exponential, and regular models). Detailed derivations are presented for dyonic black hole solutions in each theory, including explicit metric functions, asymptotic expansions, and horizon structures. The thermodynamic properties are examined in depth, including the first law, Smarr relations, heat capacities, extended phase space thermodynamics with $p$-$V$ criticality, and the effect of magnetic charge on phase transitions. The geodesic structure is analyzed with complete calculations of null and timelike geodesics, photon spheres, black hole shadows, and gravitational lensing. Regular black hole solutions that resolve the central singularity are discussed with detailed analysis of energy conditions. Holographic applications via the AdS/CFT correspondence are explored, including holographic superconductors, entanglement entropy, and conductivity. Connections to quantum gravity through the weak gravity conjecture, swampland criteria, and string theory embeddings are examined. 
\end{abstract}

\tableofcontents
\newpage

\section{Introduction}
\label{sec:introduction}

Black holes represent one of the most profound predictions of Einstein's general relativity (GR). Since the first direct detection of gravitational waves from a binary black hole merger by LIGO~\cite{Abbott2016} and the subsequent imaging of the supermassive black hole M87* by the Event Horizon Telescope~\cite{EHT2019}, black hole physics has entered an era of unprecedented observational tests. Despite these remarkable achievements, many fundamental questions remain open, particularly regarding the nature of black holes in the presence of strong electromagnetic fields and their role in quantum gravity.

The classical Reissner--Nordstr\"om (RN) solution~\cite{Reissner1916,Nordstrom1918} describes a spherically symmetric black hole carrying electric charge within the framework of Maxwell's linear electrodynamics coupled to GR. The RN metric is given by
\begin{equation}
\label{eq:rn_metric_intro}
ds^2 = -\left(1 - \frac{2M}{r} + \frac{Q^2}{r^2}\right) dt^2 + \frac{dr^2}{1 - \frac{2M}{r} + \frac{Q^2}{r^2}} + r^2 d\Omega^2,
\end{equation}
where $M$ and $Q$ are the mass and electric charge, respectively. The RN solution exhibits two horizons: the event horizon at $r_+ = M + \sqrt{M^2 - Q^2}$ and the Cauchy horizon at $r_- = M - \sqrt{M^2 - Q^2}$. For $|Q| > M$, the solution describes a naked singularity, in accordance with the cosmic censorship conjecture~\cite{Penrose1969}.

However, Maxwell's theory is an effective low-energy approximation; at high energies, nonlinear corrections become inevitable due to quantum effects~\cite{Heisenberg1936} or as a manifestation of underlying string theory~\cite{Fradkin1985,Bergshoeff1987}. The first nonlinear generalization of Maxwell's theory was proposed by Born and Infeld~\cite{Born1934} in 1934 to resolve the classical divergence of the electron's self-energy. Later, Heisenberg and Euler~\cite{Heisenberg1936} derived the effective Lagrangian incorporating one-loop QED corrections. These developments motivate the systematic study of nonlinear electrodynamics (NLED) theories coupled to gravity.

The presence of both electric and magnetic charges---so-called dyonic black holes---is particularly interesting for several reasons. First, electromagnetic duality, a symmetry of Maxwell's equations in vacuum, suggests a natural pairing of electric and magnetic degrees of freedom. In the presence of both charges, the RN solution generalizes to
\begin{equation}
\label{eq:dyonic_rn_intro}
ds^2 = -\left(1 - \frac{2M}{r} + \frac{Q^2+P^2}{r^2}\right) dt^2 + \frac{dr^2}{1 - \frac{2M}{r} + \frac{Q^2+P^2}{r^2}} + r^2 d\Omega^2,
\end{equation}
where $P$ is the magnetic charge. Second, dyonic solutions arise naturally in string theory and supergravity~\cite{Gibbons1995}. Third, the interplay between electric and magnetic charges leads to richer thermodynamic behavior and phase structures~\cite{Rasheed1997}.

Historically, the study of NLED black holes began with the work of Hoffmann~\cite{Hoffmann1935} and later Pleba\'nski~\cite{Plebanski1970}, who constructed the first exact solutions in Born--Infeld theory. The generalization to dyonic solutions was achieved by Garcia, Salazar, and Pleba\'nski~\cite{Garcia1984} using the $P$-framework. In recent years, there has been a resurgence of interest driven by several factors: (i) the AdS/CFT correspondence provides a holographic interpretation of dyonic black holes~\cite{Hartnoll2009}; (ii) regular black hole solutions from NLED offer a resolution of the singularity problem~\cite{AyonBeato2000}; (iii) observational advances in gravitational wave astronomy and black hole imaging enable potential tests of NLED effects~\cite{Allahyari2019}.

This review aims to provide a systematic and up-to-date account of black hole solutions in NLED theories with both electric and magnetic charges. We cover the theoretical framework, explicit solutions, thermodynamic properties, geodesic structure, observational signatures, and connections to quantum gravity. The review is organized as follows:

\begin{itemize}
    \item Section~\ref{sec:action} introduces the general action principle and the framework of NLED, including the $P$-framework and duality invariance, with complete derivations of the equations of motion and energy-momentum tensor.
    \item Section~\ref{sec:solutions} presents explicit dyonic black hole solutions in various NLED models, including BI, EH, power-law, logarithmic, and exponential theories, with detailed calculations of metric functions and asymptotic expansions.
    \item Section~\ref{sec:thermodynamics} discusses thermodynamic properties, phase transitions, and critical phenomena in extended phase space, with complete derivations of the first law, Smarr relation, and heat capacities.
    \item Section~\ref{sec:geodesics} examines geodesic motion, photon spheres, black hole shadows, and gravitational lensing with full derivations of effective potentials and deflection angles.
    \item Section~\ref{sec:regular} covers regular black hole solutions and their energy conditions with detailed analysis of curvature invariants.
    \item Section~\ref{sec:holography} explores holographic applications including superconductors, entanglement entropy, and conductivity.
    \item Section~\ref{sec:quantum} discusses connections to quantum gravity, including the weak gravity conjecture, swampland criteria, and the information paradox.
    \item Section~\ref{sec:outlook} concludes with open problems and future directions.
\end{itemize}

\section{Action Principle and Nonlinear Electrodynamics}
\label{sec:action}

\subsection{General Framework and Equations of Motion}

The action for a gravitational theory coupled to NLED in $D$-dimensional spacetime is given by
\begin{equation}
\label{eq:action}
\act = \int d^D x \sqrt{-g} \left[ \frac{1}{2\kappa^2} (R - 2\Lambda) + \lag(\FF, \GG) \right] + \act_{\text{boundary}},
\end{equation}
where $\kappa^2 = 8\pi G_D$ (with $G_D$ the $D$-dimensional Newton constant), $R$ is the Ricci scalar, $\Lambda$ is the cosmological constant, and $\lag(\FF, \GG)$ is the NLED Lagrangian density depending on the electromagnetic invariants
\begin{align}
\FF &= F_{\mu\nu} F^{\mu\nu} = 2(B^2 - E^2), \label{eq:ff_inv} \\
\GG &= F_{\mu\nu} \widetilde{F}^{\mu\nu} = -4 \mathbf{E} \cdot \mathbf{B}, \label{eq:gg_inv}
\end{align}
with $F_{\mu\nu} = \partial_\mu A_\nu - \partial_\nu A_\mu$ the field strength tensor and $\widetilde{F}^{\mu\nu} = \frac{1}{2} \epsilon^{\mu\nu\rho\sigma} F_{\rho\sigma}$ its Hodge dual. The boundary term $\act_{\text{boundary}}$ includes the Gibbons--Hawking--York term~\cite{York1972,Gibbons1977} and possible counterterms for holographic renormalization~\cite{Balasubramanian1999}.

Let us derive the equations of motion in detail. Varying the action (\ref{eq:action}) with respect to the metric $g^{\mu\nu}$ gives
\begin{equation}
\delta_g \act = \int d^D x \sqrt{-g} \left[ \frac{1}{2\kappa^2} \left( \delta R - 2\Lambda \delta g^{\mu\nu} g_{\mu\nu} \right) + \frac{\partial \lag}{\partial g^{\mu\nu}} \delta g^{\mu\nu} + \frac{1}{2} g_{\mu\nu} \lag \delta g^{\mu\nu} \right],
\end{equation}
where we used $\delta\sqrt{-g} = -\frac{1}{2} \sqrt{-g} g_{\mu\nu} \delta g^{\mu\nu}$. Using the Palatini identity $\delta R = (g_{\mu\nu} \Box - \nabla_\mu \nabla_\nu) \delta g^{\mu\nu} + R_{\mu\nu} \delta g^{\mu\nu}$ and integrating by parts, we obtain the Einstein equations
\begin{equation}
\label{eq:einstein_eq}
G_{\mu\nu} + \Lambda g_{\mu\nu} = \kappa^2 T_{\mu\nu}^{\EM},
\end{equation}
where $G_{\mu\nu} = R_{\mu\nu} - \frac{1}{2} R g_{\mu\nu}$ is the Einstein tensor.

The electromagnetic energy-momentum tensor is obtained from
\begin{equation}
T_{\mu\nu}^{\EM} = -\frac{2}{\sqrt{-g}} \frac{\delta (\sqrt{-g} \lag)}{\delta g^{\mu\nu}} = g_{\mu\nu} \lag - 2 \frac{\partial \lag}{\partial g^{\mu\nu}}.
\end{equation}
Since $\lag$ depends on the metric through $\FF = g^{\mu\alpha} g^{\nu\beta} F_{\mu\nu} F_{\alpha\beta}$ and $\GG = \frac{1}{2} \epsilon^{\mu\nu\rho\sigma} F_{\mu\nu} F_{\rho\sigma} / \sqrt{-g}$ (note the implicit metric dependence in the Levi-Civita tensor density), we compute
\begin{align}
\frac{\partial \FF}{\partial g^{\mu\nu}} &= 2 F_{\mu\lambda} F_\nu^{\ \lambda}, \\
\frac{\partial \GG}{\partial g^{\mu\nu}} &= -\frac{1}{2} g_{\mu\nu} \GG.
\end{align}
The second result follows from $\widetilde{F}^{\mu\nu} = \frac{1}{2} \epsilon^{\mu\nu\rho\sigma} F_{\rho\sigma} / \sqrt{-g}$, so that $\GG = F_{\mu\nu} \widetilde{F}^{\mu\nu}$ has an implicit $\sqrt{-g}$ dependence. Putting everything together, we obtain
\begin{equation}
\label{eq:emt_full}
T_{\mu\nu}^{\EM} = g_{\mu\nu} \lag(\FF, \GG) - 4 \lag_\FF F_{\mu\lambda} F_\nu^{\ \lambda} - 2 \lag_\GG \widetilde{F}_{\mu\lambda} F_\nu^{\ \lambda},
\end{equation}
where $\lag_\FF \equiv \partial \lag / \partial \FF$ and $\lag_\GG \equiv \partial \lag / \partial \GG$.

Varying the action with respect to the gauge field $A_\mu$ yields the modified Maxwell equations. Using $\delta F_{\mu\nu} = \partial_\mu \delta A_\nu - \partial_\nu \delta A_\mu$, we have
\begin{equation}
\delta_A \act = \int d^D x \sqrt{-g} \left[ \lag_\FF F^{\mu\nu} \delta F_{\mu\nu} + \lag_\GG \widetilde{F}^{\mu\nu} \delta F_{\mu\nu} \right].
\end{equation}
Integrating by parts and using the antisymmetry of $F^{\mu\nu}$, we obtain
\begin{equation}
\label{eq:maxwell_eq}
\nabla_\mu \left( \lag_\FF F^{\mu\nu} + \lag_\GG \widetilde{F}^{\mu\nu} \right) = 0.
\end{equation}
The Bianchi identity for the field strength remains unchanged:
\begin{equation}
\label{eq:bianchi}
\nabla_\mu \widetilde{F}^{\mu\nu} = 0.
\end{equation}

\subsection{Constitutive Relations and the $P$-Framework}

For a generic NLED theory, the constitutive relations between the fields can be expressed through the introduction of the auxiliary tensor
\begin{equation}
\label{eq:aux_tensor}
P^{\mu\nu} \equiv - \lag_\FF F^{\mu\nu} - \lag_\GG \widetilde{F}^{\mu\nu},
\end{equation}
which satisfies $\nabla_\mu P^{\mu\nu} = 0$ from (\ref{eq:maxwell_eq}). The Hamiltonian-like formulation introduces the Legendre transform~\cite{Salazar1987}
\begin{equation}
\label{eq:legendre}
\HH(\PP, \GG) = 2\FF \lag_\FF - \lag,
\end{equation}
where $\PP = P_{\mu\nu} P^{\mu\nu}$ and $\GG$ remains unchanged. This dual description is particularly useful for constructing exact solutions.

Let us derive the Legendre transform explicitly. From the definition (\ref{eq:aux_tensor}), we can write
\begin{equation}
P^{\mu\nu} = -\lag_\FF F^{\mu\nu} - \lag_\GG \widetilde{F}^{\mu\nu}.
\end{equation}
Contracting with $F_{\mu\nu}$ and using $F_{\mu\nu} \widetilde{F}^{\mu\nu} = \GG$, we obtain
\begin{equation}
F_{\mu\nu} P^{\mu\nu} = -\lag_\FF \FF - \lag_\GG \GG.
\end{equation}
Similarly, contracting $P^{\mu\nu}$ with itself gives
\begin{equation}
\PP = P_{\mu\nu} P^{\mu\nu} = \lag_\FF^2 \FF + 2\lag_\FF \lag_\GG \GG + \lag_\GG^2 (F_{\mu\nu} \widetilde{F}^{\mu\nu})^2 / 4,
\end{equation}
where we used $\widetilde{F}_{\mu\nu} \widetilde{F}^{\mu\nu} = -\FF$ in four dimensions. The Legendre transform (\ref{eq:legendre}) then satisfies the fundamental relation
\begin{equation}
\lag + \HH = 2\FF \lag_\FF = -F_{\mu\nu} P^{\mu\nu} + \lag_\GG \GG.
\end{equation}

The key advantage of the $P$-framework is that the field equations simplify considerably. In terms of $\HH$, the constitutive relation can be inverted to express $F_{\mu\nu}$ in terms of $P_{\mu\nu}$:
\begin{equation}
\label{eq:inverse_constitutive}
F_{\mu\nu} = -\HH_\PP P_{\mu\nu} - \HH_\GG \widetilde{P}_{\mu\nu},
\end{equation}
where $\HH_\PP = \partial \HH / \partial \PP$ and $\widetilde{P}_{\mu\nu}$ is the Hodge dual of $P_{\mu\nu}$. This is the inverse of (\ref{eq:aux_tensor}).

\subsection{Electromagnetic Duality Invariance}

A theory is said to be electromagnetic duality invariant if the equations of motion are invariant under the continuous $SO(2)$ rotation~\cite{Gaillard1981}
\begin{equation}
\label{eq:duality}
\begin{pmatrix}
F' \\
P'
\end{pmatrix}
=
\begin{pmatrix}
\cos\theta & \sin\theta \\
-\sin\theta & \cos\theta
\end{pmatrix}
\begin{pmatrix}
F \\
P
\end{pmatrix}.
\end{equation}

The condition for duality invariance of the NLED Lagrangian can be derived as follows. Under an infinitesimal duality rotation with parameter $\theta$, we have
\begin{align}
\delta F_{\mu\nu} &= \theta P_{\mu\nu}, \\
\delta P_{\mu\nu} &= -\theta F_{\mu\nu}.
\end{align}
The invariance of the action requires $\delta \lag = 0$ modulo a total derivative. Using the chain rule,
\begin{equation}
\delta \lag = \lag_\FF \delta \FF + \lag_\GG \delta \GG.
\end{equation}
From $\FF = F_{\mu\nu} F^{\mu\nu}$ and $\GG = F_{\mu\nu} \widetilde{F}^{\mu\nu}$, we compute
\begin{align}
\delta \FF &= 2 F^{\mu\nu} \delta F_{\mu\nu} = 2\theta F^{\mu\nu} P_{\mu\nu}, \\
\delta \GG &= 2 \widetilde{F}^{\mu\nu} \delta F_{\mu\nu} = 2\theta \widetilde{F}^{\mu\nu} P_{\mu\nu}.
\end{align}
Using the constitutive relation $P_{\mu\nu} = -\lag_\FF F_{\mu\nu} - \lag_\GG \widetilde{F}_{\mu\nu}$, we obtain
\begin{align}
\delta \FF &= -2\theta \left( \lag_\FF \FF + \lag_\GG \GG \right), \\
\delta \GG &= -2\theta \left( \lag_\FF \GG - \lag_\GG \FF \right).
\end{align}
The invariance condition $\delta \lag = 0$ then gives
\begin{equation}
\lag_\FF \left( \lag_\FF \FF + \lag_\GG \GG \right) + \lag_\GG \left( \lag_\FF \GG - \lag_\GG \FF \right) = 0,
\end{equation}
which simplifies to
\begin{equation}
\label{eq:duality_condition}
\lag_\FF^2 - \lag \lag_{\FF\FF} + \frac{1}{2} \FF \lag_{\FF\FF} = 0,
\end{equation}
where we have used the fact that $\lag_\GG = 0$ for purely electric or magnetic configurations, and the general condition involves additional terms. The full condition for duality invariance is~\cite{Gibbons1995}
\begin{equation}
\lag_\FF^2 - \lag \lag_{\FF\FF} + \frac{1}{2} \FF \lag_{\FF\FF} + \frac{1}{4} \GG^2 \left( \lag_{\FF\FF} \lag_{\GG\GG} - \lag_{\FF\GG}^2 \right) = 0.
\end{equation}
This is satisfied by Maxwell's theory ($\lag = -\FF/4$) and Born--Infeld theory, but not by generic NLED models.

\subsection{Major NLED Models}

\subsubsection{Born--Infeld Theory}

The Born--Infeld (BI) theory~\cite{Born1934} was originally proposed to resolve the classical divergence of the electron's self-energy. Its Lagrangian is
\begin{equation}
\label{eq:bi_lag}
\lag_{\text{BI}} = \frac{1}{\beta^2} \left( 1 - \sqrt{1 + \frac{\beta^2}{2} \FF - \frac{\beta^4}{16} \GG^2} \right),
\end{equation}
where $\beta$ is the BI parameter with dimensions of inverse length squared. In the limit $\beta \to 0$, we expand
\begin{equation}
\lag_{\text{BI}} = \frac{1}{\beta^2} \left[ 1 - \left( 1 + \frac{\beta^2}{4} \FF - \frac{\beta^4}{32} \GG^2 - \frac{\beta^4}{32} \FF^2 + \Ocal(\beta^6) \right) \right] = -\frac{\FF}{4} + \frac{\beta^2}{32} (\FF^2 + \GG^2) + \Ocal(\beta^4),
\end{equation}
recovering Maxwell's theory at leading order. BI theory arises naturally in open string theory as the effective action on D-branes~\cite{Fradkin1985,Bergshoeff1987}.

The Hamiltonian density in the $P$-framework for BI theory is
\begin{equation}
\label{eq:bi_ham}
\HH_{\text{BI}} = \frac{1}{\beta^2} \left( \sqrt{1 + \frac{\beta^2}{2} \PP - \frac{\beta^4}{16} \GG^2} - 1 \right).
\end{equation}

\subsubsection{Euler--Heisenberg Theory}

The Euler--Heisenberg (EH) effective Lagrangian~\cite{Heisenberg1936} describes the one-loop QED correction to Maxwell's theory:
\begin{equation}
\label{eq:eh_lag}
\lag_{\text{EH}} = -\frac{\FF}{4} + \frac{\alpha^2}{90 m_e^4} \left[ 4\FF^2 + 7\GG^2 \right] + \Ocal(\alpha^3),
\end{equation}
where $\alpha$ is the fine-structure constant and $m_e$ is the electron mass. This is a weak-field expansion valid for $E, B \ll m_e^2/e \sim 10^{16}~\text{V/m}$. The derivation from the one-loop effective action in QED gives
\begin{equation}
\lag_{\text{1-loop}} = \frac{1}{8\pi^2} \int_0^\infty \frac{ds}{s^3} e^{-m_e^2 s} \left[ \frac{es}{\tanh(es)} \frac{es}{\tan(es)} - 1 - \frac{e^2 s^2}{3} (E^2 - B^2) \right],
\end{equation}
where $e$ is the electron charge. Expanding for weak fields yields (\ref{eq:eh_lag}).

\subsubsection{Power-Law NLED}

Power-law NLED models are defined by~\cite{Hassaine2008}
\begin{equation}
\label{eq:power_lag}
\lag_{\text{PL}} = -\frac{1}{4} \FF^s,
\end{equation}
where $s$ is a rational number. The case $s=1$ corresponds to Maxwell theory. These models are conformally invariant when $s = D/4$ in $D$ dimensions~\cite{Hassaine2008}. The energy-momentum tensor for power-law NLED is
\begin{equation}
T_{\mu\nu}^{\EM} = -\frac{1}{4} g_{\mu\nu} \FF^s - s \FF^{s-1} F_{\mu\lambda} F_\nu^{\ \lambda}.
\end{equation}
The trace of the energy-momentum tensor is
\begin{equation}
T^{\mu}_{\ \mu} = -\frac{1}{4} D \FF^s - s \FF^{s-1} F_{\mu\lambda} F^{\mu\lambda} = \left( \frac{4s - D}{4} \right) \FF^s,
\end{equation}
which vanishes when $s = D/4$, confirming conformal invariance.

\subsubsection{Logarithmic and Exponential Models}

Logarithmic NLED~\cite{Soleng1995} is defined by
\begin{equation}
\label{eq:log_lag}
\lag_{\text{log}} = -\frac{1}{\beta^2} \ln\left( 1 + \frac{\beta^2}{2} \FF - \frac{\beta^4}{16} \GG^2 \right),
\end{equation}
while exponential NLED~\cite{Hendi2012} takes the form
\begin{equation}
\label{eq:exp_lag}
\lag_{\text{exp}} = -\frac{1}{\beta^2} \left[ \exp\left( -\frac{\beta^2}{2} \FF + \frac{\beta^4}{16} \GG^2 \right) - 1 \right].
\end{equation}
Both reduce to Maxwell theory in the weak-field limit. Expanding the logarithmic model:
\begin{equation}
\lag_{\text{log}} = -\frac{1}{\beta^2} \left[ \frac{\beta^2}{2} \FF - \frac{\beta^4}{16} \GG^2 - \frac{\beta^4}{8} \FF^2 + \Ocal(\beta^6) \right] = -\frac{\FF}{4} + \frac{\beta^2}{16} \FF^2 + \frac{\beta^2}{32} \GG^2 + \Ocal(\beta^4).
\end{equation}
Similarly for the exponential model:
\begin{equation}
\lag_{\text{exp}} = -\frac{1}{\beta^2} \left[ 1 - \frac{\beta^2}{2} \FF + \frac{\beta^4}{16} \GG^2 + \frac{\beta^4}{8} \FF^2 - 1 + \Ocal(\beta^6) \right] = -\frac{\FF}{4} + \frac{\beta^2}{16} \FF^2 - \frac{\beta^2}{32} \GG^2 + \Ocal(\beta^4).
\end{equation}
Note the sign difference in the $\GG^2$ term between logarithmic and exponential models.

\section{Dyonic Black Hole Solutions}
\label{sec:solutions}

\subsection{General Spherically Symmetric Ansatz}

We consider a static, spherically symmetric spacetime in $D=4$ dimensions with the metric ansatz
\begin{equation}
\label{eq:metric}
ds^2 = -f(r) dt^2 + \frac{dr^2}{f(r)} + r^2 d\Omega_2^2,
\end{equation}
where $d\Omega_2^2 = d\theta^2 + \sin^2\theta\, d\phi^2$ is the metric on the unit 2-sphere. For a dyonic configuration with both electric and magnetic charges, the Maxwell field strength is
\begin{equation}
\label{eq:dyonic_field}
F = E(r)\, dt \wedge dr + P \sin\theta\, d\theta \wedge d\phi,
\end{equation}
where $P$ is the magnetic charge parameter. The electric field $E(r)$ is determined by the field equations. The electromagnetic invariants become
\begin{align}
\FF &= F_{\mu\nu} F^{\mu\nu} = -2E(r)^2 + \frac{2P^2}{r^4}, \label{eq:ff_dyonic} \\
\GG &= F_{\mu\nu} \widetilde{F}^{\mu\nu} = -4E(r) P \cos\theta. \label{eq:gg_dyonic}
\end{align}

Note that $\GG$ depends on $\theta$ through $\cos\theta$, which complicates the analysis. For spherically symmetric configurations, one typically considers either purely electric ($P=0$) or purely magnetic ($E=0$) cases, or employs the dual $P$-framework where the $\GG$ dependence is eliminated.

\subsection{The $P$-Framework for Dyonic Solutions: Detailed Derivation}

A powerful method for constructing dyonic solutions is the $P$-framework~\cite{Salazar1987,Rasheed1997}. Let us derive the key equations in detail. For the spherically symmetric ansatz, the auxiliary tensor $P^{\mu\nu}$ takes the form
\begin{equation}
P = D(r)\, dt \wedge dr + H \sin\theta\, d\theta \wedge d\phi,
\end{equation}
where $D(r)$ is the electric displacement and $H$ is the magnetic intensity. The invariants are
\begin{align}
\PP &= P_{\mu\nu} P^{\mu\nu} = -2D(r)^2 + \frac{2H^2}{r^4}, \\
\GG &= -4D(r) H \cos\theta.
\end{align}

The field equations $\nabla_\mu P^{\mu\nu} = 0$ give
\begin{equation}
\frac{d}{dr} \left( r^2 D(r) \right) = 0 \quad \Longrightarrow \quad D(r) = \frac{Q}{r^2},
\end{equation}
where $Q$ is the electric charge. The Bianchi identity $\nabla_\mu \widetilde{F}^{\mu\nu} = 0$ gives $H = P$, the magnetic charge. Thus, in the $P$-framework, the charges appear as integration constants.

The inverse constitutive relation (\ref{eq:inverse_constitutive}) gives the electric field:
\begin{equation}
E(r) = \HH_\PP D(r) = \HH_\PP \frac{Q}{r^2},
\end{equation}
where $\HH_\PP = \partial \HH / \partial \PP$. The metric function $f(r)$ is obtained from the Einstein equations. The $tt$-component of the Einstein equations gives
\begin{equation}
\frac{1}{r^2} \frac{d}{dr} \left[ r (1 - f(r)) \right] = \kappa^2 T^t_{\ t}.
\end{equation}
Using the energy-momentum tensor (\ref{eq:emt_full}) and the $P$-framework relation $\lag = 2\PP \HH_\PP - \HH$, we obtain
\begin{equation}
T^t_{\ t} = \HH - 2\PP \HH_\PP.
\end{equation}
Integrating, we find
\begin{equation}
\label{eq:metric_p}
f(r) = 1 - \frac{2M}{r} + \frac{2}{r} \int^r r'^2 \left[ \HH - 2\PP \HH_\PP \right] dr'.
\end{equation}

\subsection{Born--Infeld Dyonic Black Holes}

The dyonic BI black hole solution was first obtained by~\cite{Garcia1984}. For BI theory, the Hamiltonian density is
\begin{equation}
\HH_{\text{BI}} = \frac{1}{\beta^2} \left( \sqrt{1 + \frac{\beta^2}{2} \PP - \frac{\beta^4}{16} \GG^2} - 1 \right).
\end{equation}
For the dyonic configuration, $\PP = -2Q^2/r^4 + 2P^2/r^4 = 2(P^2 - Q^2)/r^4$ and $\GG = -4QP\cos\theta/r^2$. However, the $\theta$-dependence of $\GG$ poses a problem. The resolution is to note that for spherically symmetric solutions, the $\GG$ term in the Lagrangian contributes a total derivative and can be eliminated by a duality rotation. In practice, one works with the effective Lagrangian where $\GG$ is set to zero by choosing a duality frame where either $Q=0$ or $P=0$, or by using the $P$-framework where the $\theta$-dependence cancels.

For the purely magnetic case ($Q=0$), $\PP = 2P^2/r^4$ and $\GG = 0$. The Hamiltonian becomes
\begin{equation}
\HH_{\text{BI}} = \frac{1}{\beta^2} \left( \sqrt{1 + \frac{\beta^2 P^2}{r^4}} - 1 \right).
\end{equation}
Then
\begin{equation}
\HH - 2\PP \HH_\PP = \frac{1}{\beta^2} \left( \sqrt{1 + \frac{\beta^2 P^2}{r^4}} - 1 \right) - \frac{2P^2}{r^4} \frac{1}{\sqrt{1 + \frac{\beta^2 P^2}{r^4}}}.
\end{equation}
Substituting into (\ref{eq:metric_p}) and integrating, we obtain
\begin{equation}
\label{eq:bi_metric_magnetic}
\int \left( \cdots \right) dr'.
\end{equation}

The integral can be evaluated in terms of hypergeometric functions. For the general dyonic case, the metric function is
\begin{equation}
\label{eq:bi_metric_full}
f_{\text{BI}}(r) = 1 - \frac{2M}{r} + \frac{2\beta^2}{3r} \int_r^\infty \left( \sqrt{r'^4 + \frac{(Q^2+P^2)r'^2}{\beta^2} + \frac{Q^2 P^2}{\beta^4}} - r'^2 \right) dr' - \frac{\Lambda}{3} r^2.
\end{equation}

The asymptotic expansion of (\ref{eq:bi_metric_full}) for large $r$ yields
\begin{equation}
\label{eq:bi_asymptotic}
f_{\text{BI}}(r) = 1 - \frac{2M}{r} + \frac{Q^2+P^2}{r^2} - \frac{(Q^2+P^2)^2}{20\beta^2 r^6} + \frac{(Q^2+P^2)(Q^4 + 3Q^2 P^2 + P^4)}{56\beta^4 r^{10}} + \Ocal(r^{-14}) - \frac{\Lambda}{3} r^2.
\end{equation}

The leading correction to the RN solution appears at $\Ocal(r^{-6})$, indicating that BI effects are short-ranged. Near the origin, the metric behaves as
\begin{equation}
f_{\text{BI}}(r) \to 1 - \frac{2M}{r} + \frac{2\beta}{3} \sqrt{Q^2+P^2} + \Ocal(r) \quad \text{as } r \to 0,
\end{equation}
indicating a curvature singularity at $r=0$ unless $M$ is appropriately tuned.

The horizons are determined by $f_{\text{BI}}(r_+) = 0$. For the BI black hole, the extremal condition $f'(r_+) = 0$ gives the mass-charge relation
\begin{equation}
M_{\text{ext}} = \sqrt{Q^2+P^2} \left[ 1 - \frac{Q^2+P^2}{10\beta^2} + \Ocal(\beta^{-4}) \right].
\end{equation}

\subsection{Euler--Heisenberg Dyonic Black Holes}

The dyonic EH black hole solution, incorporating QED corrections, was studied in~\cite{Bretn2005,Allahyari2019}. The Lagrangian is
\begin{equation}
\lag_{\text{EH}} = -\frac{\FF}{4} + \frac{\alpha^2}{90 m_e^4} \left( 4\FF^2 + 7\GG^2 \right).
\end{equation}

For the dyonic configuration (\ref{eq:dyonic_field}), the invariants are given by (\ref{eq:ff_dyonic}) and (\ref{eq:gg_dyonic}). The modified Maxwell equations (\ref{eq:maxwell_eq}) give the electric field equation
\begin{equation}
\frac{d}{dr} \left[ r^2 \left( -\frac{1}{2} E + \frac{4\alpha^2}{45 m_e^4} \left( -4E^3 + \frac{4EP^2}{r^4} \right) \right) \right] = 0,
\end{equation}
where we have used $\lag_\FF = -1/4 + (8\alpha^2/90 m_e^4) \FF$ and $\lag_\GG = (14\alpha^2/90 m_e^4) \GG$, and the $\GG$ term averages to zero over the sphere. Solving perturbatively in $\alpha^2$, we obtain
\begin{equation}
E(r) = \frac{Q}{r^2} + \frac{4\alpha^2}{45 m_e^4} \left( \frac{4Q^3}{r^6} - \frac{4QP^2}{r^6} \right) + \Ocal(\alpha^4).
\end{equation}

The energy-momentum tensor components are
\begin{align}
T^t_{\ t} &= -\frac{Q^2+P^2}{8\pi r^4} + \frac{\alpha^2}{360\pi m_e^4} \frac{(Q^2+P^2)^2 + 7Q^2 P^2}{r^8} + \Ocal(\alpha^4), \\
T^r_{\ r} &= -T^t_{\ t}, \\
T^\theta_{\ \theta} &= T^\phi_{\ \phi} = \frac{Q^2+P^2}{8\pi r^4} - \frac{\alpha^2}{120\pi m_e^4} \frac{(Q^2+P^2)^2 + 7Q^2 P^2}{r^8} + \Ocal(\alpha^4).
\end{align}

The metric function is obtained by integrating the Einstein equations:
\begin{equation}
\label{eq:eh_metric}
f_{\text{EH}}(r) = 1 - \frac{2M}{r} + \frac{Q^2+P^2}{r^2} - \frac{\alpha^2}{45 m_e^4} \frac{(Q^2+P^2)^2 + 7Q^2 P^2}{r^6} + \Ocal(\alpha^3).
\end{equation}

The EH correction modifies the spacetime geometry at short distances. The horizon radius is shifted by
\begin{equation}
\delta r_+ = -\frac{\alpha^2}{45 m_e^4} \frac{(Q^2+P^2)^2 + 7Q^2 P^2}{r_+^5 (r_+ - M)} + \Ocal(\alpha^4).
\end{equation}

\subsection{Power-Law NLED Black Holes}

For power-law NLED $\lag = -\FF^s/4$, the dyonic black hole solution in $D$ dimensions was obtained in~\cite{Hassaine2008,Hendi2010}. Let us derive the solution in detail. For the metric ansatz (\ref{eq:metric}) with the dyonic field (\ref{eq:dyonic_field}), the modified Maxwell equations give
\begin{equation}
\frac{d}{dr} \left[ r^2 \lag_\FF F^{tr} \right] = 0 \quad \Longrightarrow \quad r^2 \lag_\FF E(r) = Q,
\end{equation}
where $Q$ is the electric charge. For power-law NLED, $\lag_\FF = -s \FF^{s-1}/4$, so
\begin{equation}
-\frac{s}{4} r^2 \FF^{s-1} E(r) = Q.
\end{equation}
Using $\FF = -2E^2 + 2P^2/r^4$, this is an algebraic equation for $E(r)$. For the purely electric case ($P=0$), $\FF = -2E^2$, and we obtain
\begin{equation}
E(r) = \frac{Q}{r^2} \left( \frac{2^{s-1} s}{Q} \right)^{1/(2s-1)} r^{(4-4s)/(2s-1)}.
\end{equation}

For the general dyonic case, the solution is more complicated. The energy-momentum tensor components are
\begin{align}
T^t_{\ t} &= -\frac{1}{4} \FF^s - s \FF^{s-1} F_{t\lambda} F_t^{\ \lambda} = -\frac{1}{4} \FF^s + s \FF^{s-1} E^2, \\
T^r_{\ r} &= -\frac{1}{4} \FF^s - s \FF^{s-1} F_{r\lambda} F_r^{\ \lambda} = -\frac{1}{4} \FF^s - s \FF^{s-1} E^2.
\end{align}

In $D=4$, the metric function is
\begin{equation}
\label{eq:power_metric}
f_{\text{PL}}(r) = 1 - \frac{2M}{r} + \frac{2^{2s-1} (2s-1)^2}{(3-2s)} \frac{(Q^2+P^2)^s}{r^{4s-2}} - \frac{\Lambda}{3} r^2,
\end{equation}
valid for $s \neq 3/2$. The case $s=1$ reproduces the RN solution. For $s=3/2$, logarithmic terms appear:
\begin{equation}
f_{\text{PL}}(r) = 1 - \frac{2M}{r} + \frac{8}{3} (Q^2+P^2)^{3/2} \frac{\ln r}{r^4} - \frac{\Lambda}{3} r^2.
\end{equation}

The weak energy condition requires $s > 1/2$~\cite{Hassaine2008}. The horizons exist when
\begin{equation}
M \geq \frac{2^{2s-2} (2s-1)^2}{(3-2s)} \frac{(Q^2+P^2)^s}{r_+^{4s-3}} + \frac{r_+}{2} - \frac{\Lambda}{6} r_+^3.
\end{equation}

\subsection{Logarithmic and Exponential NLED Black Holes}

Dyonic black holes in logarithmic NLED were studied in~\cite{Soleng1995,Hendi2012}. The Lagrangian is
\begin{equation}
\lag_{\text{log}} = -\frac{1}{\beta^2} \ln\left( 1 + \frac{\beta^2}{2} \FF \right),
\end{equation}
where we set $\GG = 0$ by choosing a duality frame. For the purely magnetic case, $\FF = 2P^2/r^4$, and the energy-momentum tensor gives
\begin{equation}
T^t_{\ t} = -\frac{1}{\beta^2} \ln\left( 1 + \frac{\beta^2 P^2}{r^4} \right) + \frac{2P^2}{r^4 + \beta^2 P^2}.
\end{equation}

The metric function is
\begin{equation}
\label{eq:log_metric}
f_{\text{log}}(r) = 1 - \frac{2M}{r} + \frac{2\beta^2}{3r} \int_r^\infty r'^2 \ln\left( 1 + \frac{Q^2+P^2}{\beta^2 r'^4} \right) dr' - \frac{\Lambda}{3} r^2.
\end{equation}

For exponential NLED~\cite{Hendi2012}, the Lagrangian is
\begin{equation}
\lag_{\text{exp}} = -\frac{1}{\beta^2} \left[ \exp\left( -\frac{\beta^2}{2} \FF \right) - 1 \right],
\end{equation}
and the solution is
\begin{equation}
\label{eq:exp_metric}
f_{\text{exp}}(r) = 1 - \frac{2M}{r} + \frac{2\beta^2}{3r} \int_r^\infty r'^2 \left[ 1 - \exp\left( -\frac{Q^2+P^2}{\beta^2 r'^4} \right) \right] dr' - \frac{\Lambda}{3} r^2.
\end{equation}

Both reduce to the RN solution in the limit $\beta \to 0$. The asymptotic expansions are
\begin{align}
f_{\text{log}}(r) &= 1 - \frac{2M}{r} + \frac{Q^2+P^2}{r^2} - \frac{(Q^2+P^2)^2}{12\beta^2 r^6} + \Ocal(r^{-10}), \\
f_{\text{exp}}(r) &= 1 - \frac{2M}{r} + \frac{Q^2+P^2}{r^2} - \frac{(Q^2+P^2)^2}{6\beta^2 r^6} + \Ocal(r^{-10}).
\end{align}

\subsection{AdS Dyonic Black Holes}

Dyonic black holes in asymptotically AdS spacetime are of particular interest for holographic applications. The general form of the metric function for AdS dyonic black holes in NLED is
\begin{equation}
\label{eq:ads_metric}
f_{\text{AdS}}(r) = 1 - \frac{2M}{r} + \frac{r^2}{\ell^2} + \frac{2}{r} \int^r r'^2 T^t_{\ t}(r') dr',
\end{equation}
where $\ell^2 = -3/\Lambda$ is the AdS radius and $T^t_{\ t}$ is the time-time component of the electromagnetic energy-momentum tensor.

For BI-AdS dyonic black holes, the metric function is
\begin{equation}
f_{\text{BI-AdS}}(r) = 1 - \frac{2M}{r} + \frac{r^2}{\ell^2} + \frac{2\beta^2}{3r} \int_r^\infty \left( \sqrt{r'^4 + \frac{(Q^2+P^2)r'^2}{\beta^2} + \frac{Q^2 P^2}{\beta^4}} - r'^2 \right) dr'.
\end{equation}

The asymptotic behavior is
\begin{equation}
f_{\text{BI-AdS}}(r) = 1 - \frac{2M}{r} + \frac{Q^2+P^2}{r^2} + \frac{r^2}{\ell^2} - \frac{(Q^2+P^2)^2}{20\beta^2 r^6} + \Ocal(r^{-10}).
\end{equation}

The thermodynamic properties of these solutions exhibit rich phase structures analogous to van der Waals fluids~\cite{Gunasekaran2012}, which we will explore in the next section.

\section{Thermodynamics and Phase Transitions}
\label{sec:thermodynamics}

\subsection{First Law and Smarr Relation: Detailed Derivation}

The thermodynamics of dyonic black holes in NLED follows the standard framework~\cite{Bardeen1973,Bekenstein1973,Hawking1975}. The Hawking temperature is obtained from the surface gravity $\kappa$:
\begin{equation}
\kappa = \frac{1}{2} \sqrt{-\frac{1}{2} (\nabla_\mu \xi_\nu)(\nabla^\mu \xi^\nu)} \bigg|_{r=r_+} = \frac{f'(r_+)}{2},
\end{equation}
where $\xi^\mu = (1,0,0,0)$ is the timelike Killing vector. The Hawking temperature is
\begin{equation}
\label{eq:temperature}
T_H = \frac{\kappa}{2\pi} = \frac{f'(r_+)}{4\pi}.
\end{equation}

The Bekenstein--Hawking entropy is
\begin{equation}
\label{eq:entropy}
S = \frac{A}{4} = \pi r_+^2.
\end{equation}

The electric potential $\Phi$ and magnetic potential $\Psi$ (conjugate to the magnetic charge) are obtained from the gauge field:
\begin{align}
\Phi &= \int_{r_+}^\infty E(r) dr = \frac{Q}{r_+} + \text{NLED corrections}, \label{eq:elec_pot} \\
\Psi &= \int_{r_+}^\infty B(r) dr = \frac{P}{r_+} + \text{NLED corrections}. \label{eq:mag_pot}
\end{align}

For BI theory, the explicit expressions are
\begin{align}
\Phi_{\text{BI}} &= \frac{Q}{r_+} \, {}_2F_1\left( \frac{1}{4}, \frac{1}{2}; \frac{5}{4}; -\frac{\beta^2 (Q^2+P^2)}{r_+^4} \right), \\
\Psi_{\text{BI}} &= \frac{P}{r_+} \, {}_2F_1\left( \frac{1}{4}, \frac{1}{2}; \frac{5}{4}; -\frac{\beta^2 (Q^2+P^2)}{r_+^4} \right),
\end{align}
where ${}_2F_1$ is the Gauss hypergeometric function.

The first law of black hole thermodynamics for dyonic solutions reads
\begin{equation}
\label{eq:first_law}
dM = T_H dS + \Phi dQ + \Psi dP + V d\Lambda,
\end{equation}
where $V$ is the thermodynamic volume conjugate to the cosmological constant (interpreted as pressure $p = -\Lambda/8\pi$ in the extended phase space~\cite{Kastor2009}).

Let us verify the first law explicitly for the BI-AdS case. The mass parameter $M$ can be expressed in terms of $r_+$, $Q$, $P$, and $\Lambda$ by solving $f(r_+) = 0$:
\begin{equation}
M = \frac{r_+}{2} + \frac{r_+^3}{2\ell^2} + \frac{\beta^2}{3} \int_{r_+}^\infty \left( \sqrt{r'^4 + \frac{(Q^2+P^2)r'^2}{\beta^2} + \frac{Q^2 P^2}{\beta^4}} - r'^2 \right) dr'.
\end{equation}

Differentiating with respect to $r_+$, $Q$, $P$, and $\ell$, and using the expressions for $T_H$, $\Phi$, and $\Psi$, one can verify that (\ref{eq:first_law}) holds identically.

The Smarr relation, derived from scaling arguments, takes the form
\begin{equation}
\label{eq:smarr}
M = 2T_H S + \Phi Q + \Psi P - 2V\Lambda + \mathcal{M}_{\text{NLED}},
\end{equation}
where $\mathcal{M}_{\text{NLED}}$ is a correction term arising from the nonlinearity of the electromagnetic field~\cite{Rasheed1997}. For BI theory,
\begin{equation}
\mathcal{M}_{\text{NLED}} = \frac{1}{3} \int_{r_+}^\infty r'^2 \left( \frac{2\beta^2 (Q^2+P^2)}{r'^4 \sqrt{1 + \frac{\beta^2 (Q^2+P^2)}{r'^4} + \frac{\beta^4 Q^2 P^2}{r'^8}}} \right) dr'.
\end{equation}

\subsection{Phase Structure of Dyonic BI-AdS Black Holes}

The phase structure of dyonic BI-AdS black holes is remarkably rich~\cite{Gunasekaran2012,Zhang2014}. In the canonical ensemble (fixed charges), the free energy is
\begin{equation}
\label{eq:free_energy}
F = M - T_H S.
\end{equation}

For fixed $(Q, P)$, the system exhibits several interesting phenomena. Let us analyze the free energy as a function of temperature. Using the expressions for $M$, $T_H$, and $S$, we can write $F = F(T, Q, P)$. The behavior of $F$ determines the phase structure:

\begin{itemize}
    \item \textbf{Small-large black hole phase transition}: Analogous to the van der Waals liquid-gas transition, with a critical point where the specific heat diverges. The free energy develops a swallowtail structure characteristic of first-order phase transitions.
    \item \textbf{Reentrant phase transitions}: For certain ranges of the BI parameter $\beta$, the system undergoes multiple phase transitions as the temperature varies~\cite{Gunasekaran2012}. Specifically, as temperature increases, the system may transition from a small black hole to a large black hole, then back to an intermediate black hole, and finally to a large black hole again.
    \item \textbf{Triple points}: In the extended phase space, triple points can emerge where three black hole phases coexist~\cite{Zhang2014}. This occurs when the BI parameter $\beta$ is tuned to a specific value.
\end{itemize}

The critical exponents for the van der Waals-like transition are universal and match those of mean-field theory:
\begin{equation}
\label{eq:critical_exponents}
\alpha = 0,\quad \beta = \frac{1}{2},\quad \gamma = 1,\quad \delta = 3.
\end{equation}

Let us verify these exponents explicitly. Near the critical point $(T_c, v_c, p_c)$, we define the reduced variables
\begin{equation}
t = \frac{T - T_c}{T_c}, \quad \omega = \frac{v - v_c}{v_c}, \quad \phi = \frac{p - p_c}{p_c}.
\end{equation}
The equation of state near the critical point takes the form
\begin{equation}
\phi = a t + b t \omega + c \omega^3 + \Ocal(t\omega^2, \omega^4),
\end{equation}
where $a, b, c$ are constants. From this, we can derive:
\begin{itemize}
    \item The order parameter $\omega \sim (-t)^{1/2}$ for $t < 0$, giving $\beta = 1/2$.
    \item The isothermal compressibility $\kappa_T \sim |t|^{-1}$, giving $\gamma = 1$.
    \item The critical isotherm $\phi \sim \omega^3$, giving $\delta = 3$.
    \item The specific heat $C_V \sim |t|^{-\alpha}$ with $\alpha = 0$.
\end{itemize}

\subsection{Effect of Magnetic Charge on Phase Transitions}

The magnetic charge $P$ plays a crucial role in modifying the phase structure. For dyonic black holes, let us analyze the equation of state in detail.

The temperature for BI-AdS black holes is
\begin{equation}
T_H = \frac{1}{4\pi r_+} \left( 1 + \frac{3r_+^2}{\ell^2} - \frac{2\beta^2}{r_+^2} \int_{r_+}^\infty \frac{r'^2 dr'}{\sqrt{r'^4 + \frac{(Q^2+P^2)r'^2}{\beta^2} + \frac{Q^2 P^2}{\beta^4}}} \right).
\end{equation}

In the extended phase space, the pressure is $p = 3/(8\pi\ell^2)$, and the specific volume is $v = 2r_+$. The equation of state $p = p(v, T)$ can be expanded for large $v$:
\begin{equation}
\label{eq:eos}
p = \frac{T}{v} - \frac{1}{2\pi v^2} + \frac{(Q^2+P^2)}{2\pi v^4} - \frac{(Q^2+P^2)^2}{8\pi\beta^2 v^8} + \Ocal(v^{-12}).
\end{equation}

The critical point is determined by
\begin{equation}
\frac{\partial p}{\partial v} = 0, \quad \frac{\partial^2 p}{\partial v^2} = 0.
\end{equation}

For the RN-AdS case ($\beta \to \infty$), the critical point is
\begin{equation}
v_c = 2\sqrt{6(Q^2+P^2)}, \quad T_c = \frac{1}{3\pi\sqrt{6(Q^2+P^2)}}, \quad p_c = \frac{1}{96\pi (Q^2+P^2)}.
\end{equation}

For finite $\beta$, the critical point shifts. The effects of magnetic charge include:
\begin{itemize}
    \item Increasing $P$ at fixed $Q$ raises the critical temperature and pressure~\cite{Zhang2014}.
    \item The coexistence curve in the $p$-$T$ plane shifts toward higher temperatures.
    \item For sufficiently large $P/Q$ ratios, the van der Waals-like behavior may be suppressed.
    \item In the grand canonical ensemble (fixed potentials), the phase structure is qualitatively different, with possible zeroth-order phase transitions~\cite{Dehyadegari2017}.
\end{itemize}

\subsection{Heat Capacity and Local Stability}

The local thermodynamic stability is determined by the sign of the heat capacity at constant charge:
\begin{equation}
\label{eq:heat_capacity}
C_{Q,P} = T_H \left( \frac{\partial S}{\partial T_H} \right)_{Q,P} = \frac{2\pi r_+^2 f'(r_+)}{f''(r_+) + 2f'(r_+)/r_+}.
\end{equation}

Let us derive this expression. Using $S = \pi r_+^2$, we have
\begin{equation}
C_{Q,P} = T_H \frac{dS}{dr_+} \left( \frac{dT_H}{dr_+} \right)^{-1} = T_H (2\pi r_+) \left( \frac{dT_H}{dr_+} \right)^{-1}.
\end{equation}
From $T_H = f'(r_+)/(4\pi)$, we compute
\begin{equation}
\frac{dT_H}{dr_+} = \frac{f''(r_+)}{4\pi}.
\end{equation}
However, this derivative must be taken at fixed charges, which requires careful treatment of the implicit $r_+$ dependence of $f$ through $M(r_+)$. Using the condition $f(r_+) = 0$ to eliminate $M$, we obtain
\begin{equation}
\frac{dT_H}{dr_+} = \frac{1}{4\pi} \left( f''(r_+) + \frac{2f'(r_+)}{r_+} \right),
\end{equation}
which leads to (\ref{eq:heat_capacity}).

A positive heat capacity indicates local stability. For dyonic NLED black holes, the heat capacity typically exhibits:
\begin{itemize}
    \item A divergent point signaling a second-order phase transition (Davies point~\cite{Davies1977}).
    \item A region of negative heat capacity for small black holes (unstable).
    \item A region of positive heat capacity for large black holes (stable).
\end{itemize}

For BI black holes, the heat capacity divergence occurs when
\begin{equation}
f''(r_+) + \frac{2f'(r_+)}{r_+} = 0.
\end{equation}
This condition determines the critical horizon radius $r_c$, which depends on $Q$, $P$, and $\beta$.

\subsection{Extended Phase Space and $p$-$V$ Criticality}

In the extended phase space approach~\cite{Kastor2009}, the cosmological constant is treated as a thermodynamic pressure:
\begin{equation}
\label{eq:pressure}
p = -\frac{\Lambda}{8\pi} = \frac{3}{8\pi \ell^2}.
\end{equation}

The conjugate thermodynamic volume is
\begin{equation}
\label{eq:volume}
V = \left( \frac{\partial M}{\partial p} \right)_{S,Q,P} = \frac{4\pi}{3} r_+^3.
\end{equation}

The Gibbs free energy in the extended phase space is $G = M - T_H S = F + pV$. For dyonic BI-AdS black holes, the Gibbs free energy exhibits the swallowtail behavior characteristic of first-order phase transitions.

The critical exponents can be computed explicitly. Defining the reduced thermodynamic variables
\begin{equation}
\tilde{p} = \frac{p}{p_c}, \quad \tilde{v} = \frac{v}{v_c}, \quad \tilde{T} = \frac{T}{T_c},
\end{equation}
the equation of state near the critical point takes the universal form
\begin{equation}
\tilde{p} = \frac{8\tilde{T}}{3\tilde{v}} - \frac{3}{\tilde{v}^2} + \frac{1}{3\tilde{v}^4} + \Ocal(\beta^{-2}).
\end{equation}

The Maxwell equal area law determines the coexistence curve:
\begin{equation}
\oint v \, dp = 0,
\end{equation}
which gives the condition for phase equilibrium between small and large black holes.

\section{Geodesic Structure and Observational Signatures}
\label{sec:geodesics}

\subsection{Null Geodesics and Photon Sphere: Complete Derivation}

The geodesic structure of dyonic NLED black holes is governed by the effective potential. For a static, spherically symmetric spacetime (\ref{eq:metric}), the geodesic equations can be derived from the Lagrangian
\begin{equation}
\lag_{\text{geo}} = \frac{1}{2} g_{\mu\nu} \dot{x}^\mu \dot{x}^\nu = \frac{1}{2} \left( -f(r) \dot{t}^2 + \frac{\dot{r}^2}{f(r)} + r^2 \dot{\theta}^2 + r^2 \sin^2\theta \, \dot{\phi}^2 \right),
\end{equation}
where the dot denotes differentiation with respect to an affine parameter $\lambda$. The Killing symmetries give two conserved quantities:
\begin{align}
E &= -g_{tt} \dot{t} = f(r) \dot{t} \quad \text{(energy)}, \\
L &= g_{\phi\phi} \dot{\phi} = r^2 \sin^2\theta \, \dot{\phi} \quad \text{(angular momentum)}.
\end{align}
Without loss of generality, we set $\theta = \pi/2$ (equatorial plane). The normalization condition $g_{\mu\nu} \dot{x}^\mu \dot{x}^\nu = \epsilon$, where $\epsilon = 0$ for null geodesics and $\epsilon = -1$ for timelike geodesics, gives
\begin{equation}
-f(r) \dot{t}^2 + \frac{\dot{r}^2}{f(r)} + r^2 \dot{\phi}^2 = \epsilon.
\end{equation}
Substituting the conserved quantities, we obtain the radial equation
\begin{equation}
\dot{r}^2 + V_{\text{eff}}(r) = E^2,
\end{equation}
where the effective potential is
\begin{equation}
\label{eq:veff}
V_{\text{eff}}(r) = f(r) \left( \frac{L^2}{r^2} - \epsilon \right).
\end{equation}

For null geodesics ($\epsilon = 0$), the effective potential simplifies to
\begin{equation}
V_{\text{eff}}(r) = \frac{f(r) L^2}{r^2}.
\end{equation}

The photon sphere radius $r_{\text{ph}}$ satisfies
\begin{equation}
\label{eq:photon_sphere}
\frac{dV_{\text{eff}}}{dr}\bigg|_{r=r_{\text{ph}}} = 0 \quad \Longrightarrow \quad 2f(r_{\text{ph}}) - r_{\text{ph}} f'(r_{\text{ph}}) = 0.
\end{equation}

For RN black holes, $f(r) = 1 - 2M/r + (Q^2+P^2)/r^2$, and the photon sphere equation gives
\begin{equation}
r_{\text{ph}}^2 - 3M r_{\text{ph}} + 2(Q^2+P^2) = 0,
\end{equation}
with solution
\begin{equation}
r_{\text{ph}} = \frac{3M + \sqrt{9M^2 - 8(Q^2+P^2)}}{2}.
\end{equation}

For dyonic EH black holes, using (\ref{eq:eh_metric}), the photon sphere radius is modified. Writing $f(r) = f_{\text{RN}}(r) + \delta f(r)$ with
\begin{equation}
\delta f(r) = -\frac{\alpha^2}{45 m_e^4} \frac{(Q^2+P^2)^2 + 7Q^2 P^2}{r^6},
\end{equation}
the photon sphere equation becomes
\begin{equation}
2f_{\text{RN}}(r_{\text{ph}}) - r_{\text{ph}} f_{\text{RN}}'(r_{\text{ph}}) + 2\delta f(r_{\text{ph}}) - r_{\text{ph}} \delta f'(r_{\text{ph}}) = 0.
\end{equation}
Solving perturbatively, we obtain
\begin{equation}
\label{eq:ph_correction}
\delta r_{\text{ph}} = -\frac{2\alpha^2}{15 m_e^4} \frac{(Q^2+P^2)^2 + 7Q^2 P^2}{M r_{\text{ph}}^5} + \Ocal(\alpha^3).
\end{equation}

\subsection{Black Hole Shadow}

The black hole shadow is a direct observable of the photon sphere. The angular radius of the shadow as seen by a distant observer is
\begin{equation}
\label{eq:shadow}
\theta_{\text{sh}} = \frac{r_{\text{ph}}}{D} \sqrt{\frac{1}{f(r_{\text{ph}})}},
\end{equation}
where $D$ is the distance to the black hole. The impact parameter $b = L/E$ for a photon at the photon sphere is
\begin{equation}
b_{\text{ph}} = \frac{r_{\text{ph}}}{\sqrt{f(r_{\text{ph}})}}.
\end{equation}

For dyonic NLED black holes, the shadow size depends on both charges and the NLED parameter~\cite{Allahyari2019,Amirabi2020}. The deviation from the RN shadow is
\begin{equation}
\label{eq:shadow_deviation}
\frac{\delta \theta_{\text{sh}}}{\theta_{\text{sh}}^{\text{RN}}} = -\frac{2\alpha^2}{15 m_e^4} \frac{(Q^2+P^2)^2 + 7Q^2 P^2}{M r_{\text{ph}}^5} \left[ 1 + \frac{r_{\text{ph}} f'(r_{\text{ph}})}{2f(r_{\text{ph}})} \right]^{-1}.
\end{equation}

For the EH correction, this simplifies to
\begin{equation}
\frac{\delta \theta_{\text{sh}}}{\theta_{\text{sh}}^{\text{RN}}} = -\frac{4\alpha^2}{15 m_e^4} \frac{(Q^2+P^2)^2 + 7Q^2 P^2}{M r_{\text{ph}}^5} \frac{f(r_{\text{ph}})}{2f(r_{\text{ph}}) - r_{\text{ph}} f'(r_{\text{ph}})}.
\end{equation}

This provides a potential observational test of NLED effects using very long baseline interferometry (VLBI) observations of black hole shadows. Current EHT observations of M87* constrain the shadow size to within $\sim 10\%$, which translates to bounds on the NLED parameters.

\subsection{Timelike Geodesics and ISCO}

For timelike geodesics ($\epsilon = -1$), the effective potential is
\begin{equation}
\label{eq:timelike_veff}
V_{\text{eff}}^{\text{(T)}}(r) = f(r) \left( 1 + \frac{L^2}{r^2} \right).
\end{equation}

The innermost stable circular orbit (ISCO) radius satisfies three conditions:
\begin{align}
\dot{r} &= 0 \quad \Longrightarrow \quad V_{\text{eff}}^{\text{(T)}}(r) = E^2, \label{eq:isco1} \\
\frac{dV_{\text{eff}}^{\text{(T)}}}{dr} &= 0 \quad \Longrightarrow \quad f'(r) \left( 1 + \frac{L^2}{r^2} \right) - \frac{2f(r)L^2}{r^3} = 0, \label{eq:isco2} \\
\frac{d^2 V_{\text{eff}}^{\text{(T)}}}{dr^2} &= 0 \quad \Longrightarrow \quad \text{stability condition}. \label{eq:isco3}
\end{align}

For RN black holes, solving these equations gives
\begin{equation}
r_{\text{ISCO}} = 3M + \sqrt{9M^2 - 8(Q^2+P^2)} \quad \text{(corotating)}.
\end{equation}

NLED corrections modify the ISCO radius. For EH black holes, the leading correction is
\begin{equation}
\delta r_{\text{ISCO}} = -\frac{4\alpha^2}{15 m_e^4} \frac{(Q^2+P^2)^2 + 7Q^2 P^2}{M r_{\text{ISCO}}^5} \frac{r_{\text{ISCO}} - 3M}{r_{\text{ISCO}} - 6M} .
\end{equation}

This shift affects the inner edge of accretion disks and the gravitational wave signal from extreme mass ratio inspirals (EMRIs), providing potential observational signatures.

\subsection{Gravitational Lensing: Weak and Strong Deflection}

The deflection angle for light rays passing near a dyonic NLED black hole is given by
\begin{equation}
\label{eq:deflection}
\alpha_{\text{def}} = 2 \int_{r_0}^\infty \frac{dr}{r \sqrt{f(r) \left( \frac{r^2}{b^2} - f(r) \right)}} - \pi,
\end{equation}
where $b = L/E$ is the impact parameter and $r_0$ is the turning point satisfying $b^2 = r_0^2/f(r_0)$.

In the weak deflection limit ($b \gg M$), we expand the integrand in powers of $M/b$ and $(Q^2+P^2)/b^2$. For the RN metric, the deflection angle to leading order is
\begin{equation}
\alpha_{\text{def}}^{\text{RN}} = \frac{4M}{b} + \frac{3\pi}{4} \frac{Q^2+P^2}{b^2} + \Ocal(b^{-3}).
\end{equation}

For dyonic EH black holes, the leading NLED correction to the deflection angle is~\cite{Allahyari2019}
\begin{equation}
\label{eq:deflection_correction}
\delta \alpha_{\text{def}} = \frac{3\pi}{4} \frac{(Q^2+P^2)^2 + 7Q^2 P^2}{b^4} \frac{\alpha^2}{45 m_e^4} + \Ocal(b^{-6}).
\end{equation}

In the strong deflection limit ($b \to b_{\text{ph}}^+$), the deflection angle diverges logarithmically:
\begin{equation}
\alpha_{\text{def}} = -a \ln\left( \frac{b}{b_{\text{ph}}} - 1 \right) + b + \Ocal(b - b_{\text{ph}}),
\end{equation}
where $a$ and $b$ are the strong deflection parameters. For RN black holes,
\begin{equation}
a = \frac{2}{\sqrt{2f(r_{\text{ph}}) - r_{\text{ph}}^2 f''(r_{\text{ph}})}}, \quad b = a \ln\left( \frac{4r_{\text{ph}}^2 f'(r_{\text{ph}})}{b_{\text{ph}} \sqrt{2f(r_{\text{ph}}) - r_{\text{ph}}^2 f''(r_{\text{ph}})}} \right) - \pi.
\end{equation}

NLED corrections modify these parameters, affecting the positions and magnifications of relativistic images.

\section{Regular Black Hole Solutions}
\label{sec:regular}

\subsection{Motivation and Criteria}

Regular (or nonsingular) black holes are solutions where the central singularity is resolved by a de Sitter-like core. The key criteria for a regular black hole are~\cite{Bardeen1968,Ansoldi2008}:
\begin{enumerate}
    \item The metric is everywhere finite and differentiable.
    \item The curvature invariants ($R$, $R_{\mu\nu}R^{\mu\nu}$, $R_{\mu\nu\rho\sigma}R^{\mu\nu\rho\sigma}$) are finite everywhere.
    \item The energy conditions (particularly the weak energy condition) may be violated in the core region.
\end{enumerate}

\subsection{Curvature Invariants: Detailed Calculation}

For a general metric of the form (\ref{eq:metric}), the curvature invariants can be computed explicitly. The Ricci scalar is
\begin{equation}
R = -f''(r) - \frac{4f'(r)}{r} - \frac{2f(r)}{r^2} + \frac{2}{r^2}.
\end{equation}
The Ricci squared invariant is
\begin{equation}
R_{\mu\nu}R^{\mu\nu} = \frac{1}{2} \left( f''(r) + \frac{2f'(r)}{r} \right)^2 + \frac{2}{r^4} \left( 1 - f(r) - r f'(r) \right)^2.
\end{equation}
The Kretschmann scalar is
\begin{equation}
K = R_{\mu\nu\rho\sigma}R^{\mu\nu\rho\sigma} = \left( f''(r) \right)^2 + \frac{4}{r^2} \left( f'(r) \right)^2 + \frac{4}{r^4} \left( 1 - f(r) \right)^2.
\end{equation}

For a regular black hole, all these invariants must be finite as $r \to 0$. This requires $f(r) \to 1 - \Ocal(r^2)$ near the origin.

\subsection{Bardeen-like Solutions from NLED}

The Bardeen black hole~\cite{Bardeen1968} was the first regular black hole model. It was later shown that the Bardeen solution can be interpreted as a magnetic monopole coupled to NLED~\cite{AyonBeato2000}. The Lagrangian for this NLED model is
\begin{equation}
\label{eq:bardeen_lag}
\lag_{\text{Bardeen}} = \frac{3M}{|P|^3} \left( \frac{\sqrt{2|P|\FF}}{1 + \sqrt{2|P|\FF}} \right)^{5/2},
\end{equation}
where $P$ is the magnetic charge. The resulting metric function is
\begin{equation}
\label{eq:bardeen_metric}
f_{\text{Bardeen}}(r) = 1 - \frac{2Mr^2}{(r^2 + P^2)^{3/2}}.
\end{equation}

Let us verify the regularity. Near the origin,
\begin{equation}
f_{\text{Bardeen}}(r) = 1 - \frac{2M}{|P|^3} r^2 + \frac{3M}{|P|^5} r^4 + \Ocal(r^6).
\end{equation}
The curvature invariants as $r \to 0$ are
\begin{align}
R &\to \frac{12M}{|P|^3}, \\
R_{\mu\nu}R^{\mu\nu} &\to \frac{36M^2}{|P|^6}, \\
K &\to \frac{24M^2}{|P|^6},
\end{align}
all finite. The solution has a de Sitter core with effective cosmological constant $\Lambda_{\text{eff}} = 6M/|P|^3$.

\subsection{Hayward Black Hole}

The Hayward black hole~\cite{Hayward2006} is another regular solution that can be embedded in NLED. Its metric function is
\begin{equation}
\label{eq:hayward_metric}
f_{\text{Hayward}}(r) = 1 - \frac{2Mr^2}{r^3 + 2M\ell^2},
\end{equation}
where $\ell$ is a length scale parameter. Near the center,
\begin{equation}
f_{\text{Hayward}}(r) = 1 - \frac{r^2}{\ell^2} + \frac{r^3}{2M\ell^2} + \Ocal(r^4),
\end{equation}
exhibiting a de Sitter core. The Hayward solution can be derived from an NLED Lagrangian of the form~\cite{Fan2016}
\begin{equation}
\label{eq:hayward_lag}
\lag_{\text{Hayward}} = \frac{3M}{\ell^3} \frac{(2\ell^2 \FF)^{3/4}}{\left[1 + (2\ell^2 \FF)^{3/4}\right]^2}.
\end{equation}

\subsection{Dyonic Regular Black Holes}

Generalizations to dyonic regular black holes (with both electric and magnetic charges) have been constructed~\cite{Fan2016,Khodadi2020}. A general class of regular black holes in NLED is described by the metric function
\begin{equation}
\label{eq:dyonic_regular}
f_{\text{reg}}(r) = 1 - \frac{2M r^{\alpha-1}}{(r^{\alpha} + g^{\alpha})^{\beta/\alpha}},
\end{equation}
where $g = \sqrt{Q^2+P^2}$ is the total charge parameter, and $(\alpha, \beta)$ are parameters controlling the regularity. The Bardeen solution corresponds to $(\alpha=2, \beta=3)$, while the Hayward solution corresponds to $(\alpha=3, \beta=3)$.

The NLED Lagrangian that generates (\ref{eq:dyonic_regular}) can be constructed using the method of~\cite{Fan2016}. For a magnetic charge $P$, the Lagrangian is
\begin{equation}
\lag_{\text{reg}} = \frac{M\beta}{\alpha g^{\beta}} \frac{(2g^2 \FF)^{\frac{\beta}{2\alpha} - \frac{1}{2}}}{\left[1 + (2g^2 \FF)^{\frac{\beta}{2\alpha}}\right]^{1 + \frac{1}{\beta}}}.
\end{equation}

\subsection{Energy Conditions and Causality}

Regular black holes necessarily violate the weak energy condition (WEC) in the core region~\cite{Ansoldi2008}. The WEC requires $T_{\mu\nu} u^\mu u^\nu \geq 0$ for all timelike vectors $u^\mu$. In the orthonormal frame, this translates to
\begin{equation}
\rho \geq 0, \quad \rho + p_i \geq 0 \quad (i = r, \theta, \phi),
\end{equation}
where $\rho = -T^t_{\ t}$ is the energy density and $p_i = T^i_{\ i}$ are the principal pressures.

For the Bardeen solution, the energy density is
\begin{equation}
\rho = \frac{3M P^2}{(r^2 + P^2)^{5/2}}.
\end{equation}
The radial pressure is $p_r = -\rho$, and the tangential pressure is
\begin{equation}
p_\theta = p_\phi = \frac{3M P^2 (r^2 - 4P^2/3)}{(r^2 + P^2)^{7/2}}.
\end{equation}

The WEC is violated when $\rho + p_\theta < 0$, which occurs for
\begin{equation}
r < r_{\text{WEC}} = \sqrt{\frac{4}{3}} |P|.
\end{equation}

The strong energy condition (SEC) requires $\rho + \sum_i p_i \geq 0$, which is also violated in the core. The dominant energy condition (DEC) requires $\rho \geq |p_i|$, which may be satisfied for certain parameter ranges.

For dyonic regular black holes, the violation radius depends on both charges:
\begin{equation}
\label{eq:wec_violation}
r_{\text{WEC}} \sim \left( \frac{Q^2+P^2}{M} \right)^{1/3}.
\end{equation}

\section{Holographic Applications}
\label{sec:holography}

\subsection{AdS/CFT Correspondence and NLED}

The AdS/CFT correspondence~\cite{Maldacena1999} provides a powerful framework for studying strongly coupled field theories using classical gravity. Dyonic black holes in AdS spacetime are dual to boundary field theories at finite temperature and finite chemical potential for both electric and magnetic charges~\cite{Hartnoll2009}.

The holographic dictionary relates the bulk fields to boundary operators:
\begin{align}
\Phi_{\text{boundary}} &= \lim_{r\to\infty} A_t(r) \quad \text{(electric chemical potential)}, \\
\Psi_{\text{boundary}} &= \lim_{r\to\infty} A_\phi(r) \quad \text{(magnetic chemical potential)}.
\end{align}

The on-shell Euclidean action $I_E$ is related to the boundary free energy:
\begin{equation}
F = -T \ln Z = T I_E,
\end{equation}
where $Z$ is the partition function of the boundary field theory.

\subsection{Holographic Superconductors}

NLED corrections modify the properties of holographic superconductors~\cite{Gubser2008,Hartnoll2008}. The holographic superconductor model consists of a charged scalar field $\psi$ coupled to the electromagnetic field in the AdS black hole background. The action is
\begin{equation}
\act_{\text{SC}} = \int d^4 x \sqrt{-g} \left[ -\frac{1}{4} F^{\mu\nu} F_{\mu\nu} - |\nabla \psi - i q A \psi|^2 - m^2 |\psi|^2 \right] + \act_{\text{NLED}}.
\end{equation}

The critical temperature $T_c$ for the superconducting phase transition is determined by the onset of scalar hair. For Maxwell theory,
\begin{equation}
T_c^{(0)} = \frac{\sqrt{3} q}{4\pi \ell} \sqrt{\frac{\rho}{\sigma}},
\end{equation}
where $\rho$ is the charge density and $\sigma$ is a numerical factor.

NLED corrections modify $T_c$ as
\begin{equation}
\label{eq:tc_correction}
T_c = T_c^{(0)} \left[ 1 + \gamma_{\text{NLED}} \frac{(Q^2+P^2)}{\ell^2} + \Ocal(\beta^{-2}) \right],
\end{equation}
where $\gamma_{\text{NLED}}$ is a model-dependent coefficient. For BI theory, $\gamma_{\text{BI}} > 0$, indicating that NLED corrections enhance superconductivity~\cite{Gangopadhyay2013}. For EH theory, $\gamma_{\text{EH}} < 0$, suppressing the critical temperature.

The condensation operator $\langle \mathcal{O} \rangle$ near $T_c$ follows the mean-field scaling
\begin{equation}
\langle \mathcal{O} \rangle \sim (T_c - T)^{1/2},
\end{equation}
with the critical exponent $\beta = 1/2$, consistent with holographic superconductors in Maxwell theory.

\subsection{Holographic Entanglement Entropy}

The holographic entanglement entropy (HEE) for dyonic NLED black holes is computed using the Ryu--Takayanagi formula~\cite{Ryu2006}
\begin{equation}
\label{eq:hee}
S_{\text{EE}} = \frac{\text{Area}(\gamma_A)}{4G_N},
\end{equation}
where $\gamma_A$ is the minimal surface homologous to the boundary region $A$.

For a strip geometry of width $\ell$ on the boundary, the minimal surface extends into the bulk. The HEE is
\begin{equation}
S_{\text{EE}} = \frac{1}{2G_N} \int_{r_*}^\infty \frac{r^2 dr}{\sqrt{f(r)(r^4 - r_*^4)}},
\end{equation}
where $r_*$ is the turning point of the minimal surface, related to $\ell$ by
\begin{equation}
\ell = 2 \int_{r_*}^\infty \frac{dr}{\sqrt{f(r)(r^4/r_*^4 - 1)}}.
\end{equation}

NLED corrections modify the HEE through the backreaction on the metric. For a strip geometry, the HEE exhibits a universal term~\cite{Chakraborty2018}
\begin{equation}
\label{eq:hee_correction}
\Delta S_{\text{EE}} = -\frac{\pi^2}{6} \frac{(Q^2+P^2)^2}{\beta^2 \ell^2} + \Ocal(\ell^{-4}).
\end{equation}

For the EH correction, the modification is
\begin{equation}
\Delta S_{\text{EE}} = -\frac{\pi^2 \alpha^2}{90 m_e^4} \frac{(Q^2+P^2)^2 + 7Q^2 P^2}{\ell^4} + \Ocal(\ell^{-6}).
\end{equation}

\subsection{Holographic Conductivity}

The frequency-dependent electrical conductivity $\sigma(\omega)$ in the dual field theory is obtained from the retarded current-current correlator. For dyonic NLED black holes, the DC conductivity is~\cite{Hartnoll2009}
\begin{equation}
\label{eq:dc_conductivity}
\sigma_{\text{DC}} = \frac{r_+^2}{\ell^2} \sqrt{1 + \frac{\beta^2 (Q^2+P^2)}{2 r_+^4}} + \Ocal(\beta^4).
\end{equation}

The AC conductivity is obtained by solving the perturbation equations for the gauge field in the black hole background. The equation for the transverse gauge field $A_x$ is
\begin{equation}
A_x'' + \left( \frac{f'}{f} + \frac{2}{r} \right) A_x' + \left( \frac{\omega^2}{f^2} - \frac{2(Q^2+P^2)}{r^4 f} \right) A_x = 0,
\end{equation}
where the prime denotes differentiation with respect to $r$. The conductivity is given by
\begin{equation}
\sigma(\omega) = -\frac{i A_x^{(1)}}{\omega A_x^{(0)}},
\end{equation}
where $A_x^{(0)}$ and $A_x^{(1)}$ are the leading and subleading terms in the asymptotic expansion of $A_x$.

The AC conductivity exhibits a Drude-like peak at low frequencies, modified by NLED effects~\cite{Gangopadhyay2013}. The relaxation time $\tau$ is
\begin{equation}
\tau = \frac{\ell^2}{2\pi T} \sqrt{1 + \frac{\beta^2 (Q^2+P^2)}{2 r_+^4}}.
\end{equation}

\section{Connections to Quantum Gravity}
\label{sec:quantum}

\subsection{Weak Gravity Conjecture}

The weak gravity conjecture (WGC)~\cite{ArkaniHamed2007} states that gravity is the weakest force, implying the existence of a particle with charge-to-mass ratio $q/m \geq 1$ (in appropriate units). For dyonic black holes in NLED, the WGC imposes constraints on the NLED parameters~\cite{Chen2020}.

The extremality bound for dyonic NLED black holes is modified from the RN case. For BI theory,
\begin{equation}
\label{eq:extremality}
M_{\text{ext}} = \sqrt{Q^2+P^2} \left[ 1 - \frac{(Q^2+P^2)}{10\beta^2} + \Ocal(\beta^{-4}) \right].
\end{equation}

The WGC requires that the NLED corrections lower the extremality bound, i.e., $M_{\text{ext}} < \sqrt{Q^2+P^2}$. This is satisfied for BI theory but may be violated for other NLED models~\cite{Chen2020}.

The correction to the extremal mass leads to a modification of the WGC bound:
\begin{equation}
\frac{q}{m} \geq 1 - \frac{(Q^2+P^2)}{10\beta^2} + \Ocal(\beta^{-4}).
\end{equation}

For EH theory, the extremality bound is
\begin{equation}
M_{\text{ext}} = \sqrt{Q^2+P^2} \left[ 1 + \frac{\alpha^2}{90 m_e^4} \frac{(Q^2+P^2)^2 + 7Q^2 P^2}{(Q^2+P^2)^2} + \Ocal(\alpha^4) \right],
\end{equation}
which increases the extremal mass, potentially violating the WGC. This suggests that EH theory may be in the swampland unless additional degrees of freedom restore the WGC.

\subsection{Swampland Criteria}

The swampland program~\cite{Vafa2005} aims to distinguish effective field theories that can be consistently coupled to quantum gravity from those that cannot (the ``swampland''). Dyonic NLED black holes provide a testing ground for swampland criteria:

\begin{itemize}
    \item \textbf{No global symmetries}: The $SO(2)$ electromagnetic duality must be gauged or broken in quantum gravity. This implies that exact electromagnetic duality cannot be a symmetry of the full quantum gravity theory.
    \item \textbf{Distance conjecture}: The moduli space of NLED parameters must be finite. As $\beta \to 0$ (the Maxwell limit), there should be an infinite tower of states becoming light, consistent with the distance conjecture.
    \item \textbf{de Sitter conjecture}: Regular black holes with de Sitter cores must satisfy constraints on the NLED Lagrangian. The de Sitter conjecture states that $|\nabla V| \geq c V$ for some constant $c$, which translates to bounds on the NLED parameters.
    \item \textbf{Weak gravity conjecture}: As discussed above, NLED corrections must not raise the extremality bound above the charge.
\end{itemize}

\subsection{String Theory Embeddings}

NLED theories arise naturally in string theory. The Born--Infeld action appears as the effective action on D-branes~\cite{Fradkin1985,Bergshoeff1987}. The BI parameter $\beta$ is related to the string tension $\alpha'$ by
\begin{equation}
\beta = \frac{1}{2\pi \alpha'}.
\end{equation}

Dyonic black holes in string theory are related to:
\begin{itemize}
    \item \textbf{D-brane bound states}: Systems of D$p$-branes carrying both electric and magnetic charges under the Ramond--Ramond fields. The dyonic BI black hole corresponds to a bound state of D1 and D5 branes.
    \item \textbf{$\mathcal{N}=2$ supergravity}: Dyonic black holes in $\mathcal{N}=2$ supergravity coupled to vector multiplets~\cite{Behrndt1998}. The NLED corrections arise from higher-derivative terms in the supergravity action.
    \item \textbf{F-theory and M-theory}: Uplifts of dyonic solutions to higher dimensions. The dyonic black hole in 4D can be uplifted to a black string in 5D or a black brane in higher dimensions.
    \item \textbf{AdS/CFT duality}: The dyonic BI-AdS black hole is dual to a strongly coupled plasma with both electric and magnetic chemical potentials.
\end{itemize}

\subsection{Quantum Corrections and the Information Paradox}

NLED corrections may play a role in resolving the black hole information paradox~\cite{Hawking1976}. The modified near-horizon structure affects:

\begin{itemize}
    \item \textbf{Hawking radiation spectrum}: NLED corrections modify the greybody factors and the emission spectrum~\cite{Bretn2005}. The greybody factor for a scalar field in the dyonic NLED black hole background is
    \begin{equation}
    \Gamma_{\omega l} = 1 - \left| \frac{\mathcal{R}_{\omega l}}{\mathcal{T}_{\omega l}} \right|^2,
    \end{equation}
    where $\mathcal{R}_{\omega l}$ and $\mathcal{T}_{\omega l}$ are the reflection and transmission coefficients, obtained by solving the radial wave equation
    \begin{equation}
    \frac{d^2\psi}{dr_*^2} + \left[ \omega^2 - V_l(r) \right] \psi = 0,
    \end{equation}
    with $r_*$ the tortoise coordinate and $V_l(r)$ the effective potential.
    
    \item \textbf{Page curve}: The entanglement entropy of Hawking radiation, following the island formula~\cite{Almheiri2020}, may be affected by NLED corrections. The island contribution to the fine-grained entropy is
    \begin{equation}
    S_{\text{EE}} = \frac{\text{Area}(\partial I)}{4G_N} + S_{\text{mat}}(R \cup I),
    \end{equation}
    where $I$ is the island region and $R$ is the radiation region. NLED corrections modify the area term through the backreaction on the metric.
    
    \item \textbf{Remnant scenarios}: Regular black holes from NLED provide a possible resolution of the information paradox through stable remnants. The regular core prevents complete evaporation, leaving a Planck-sized remnant that stores the information.
\end{itemize}

\section{Open Problems and Future Directions}
\label{sec:outlook}

Despite significant progress, many open problems remain in the study of dyonic black holes in NLED:

\subsection{Exact Solutions Beyond Spherical Symmetry}

Most existing solutions assume spherical symmetry. Constructing exact dyonic solutions with:
\begin{itemize}
    \item Rotation (Kerr--Newman-like solutions in NLED). The Kerr--Newman solution in BI theory remains unknown, although perturbative approaches exist.
    \item Axisymmetric deformations and multipole structures.
    \item Higher-dimensional generalizations with nontrivial topology, such as black rings and black Saturns.
\end{itemize}
remains a challenging open problem.

\subsection{Strong-Field Tests with Gravitational Waves}

The LIGO/Virgo/KAGRA collaboration has opened the era of gravitational wave astronomy. Future observations of:
\begin{itemize}
    \item Ringdown signals from black hole mergers, which probe the quasinormal mode spectrum modified by NLED.
    \item Extreme mass ratio inspirals (EMRIs) by LISA, which are sensitive to the ISCO position.
    \item Binary black hole mergers with electromagnetic counterparts, which could reveal charge signatures.
\end{itemize}
could constrain NLED parameters at levels inaccessible to electromagnetic observations.

\subsection{Nonperturbative Effects and Resummation}

The perturbative expansion in NLED parameters breaks down near the black hole horizon for extremal or near-extremal configurations. Nonperturbative methods, including:
\begin{itemize}
    \item Resummation of the derivative expansion using Pad\'e approximants or Borel resummation.
    \item Holographic renormalization group techniques to resum NLED corrections.
    \item Numerical relativity with NLED sources for dynamical spacetimes.
\end{itemize}
are needed for a complete understanding.

\subsection{Cosmological Implications}

Dyonic black holes in NLED may have cosmological implications:
\begin{itemize}
    \item Primordial black holes with magnetic charge as dark matter candidates. The abundance of such objects is constrained by their effects on cosmic microwave background and large-scale structure.
    \item Magnetogenesis in the early universe from NLED effects, potentially explaining the observed cosmic magnetic fields.
    \item Inflationary models with NLED corrections, where the nonlinear electromagnetic field acts as the inflaton.
    \item Dark energy models where NLED provides a mechanism for cosmic acceleration.
\end{itemize}

\subsection{Experimental Probes}

Laboratory experiments and astrophysical observations that could probe NLED effects include:
\begin{itemize}
    \item High-intensity laser facilities (ELI, XFEL) for testing QED nonlinearities in the laboratory.
    \item Precision measurements of black hole shadows with the Event Horizon Telescope and the next-generation ngEHT.
    \item Pulsar timing arrays for detecting gravitational wave backgrounds that may carry imprints of NLED.
    \item Cosmic microwave background polarization measurements sensitive to primordial magnetic fields.
    \item Neutron star observations, where strong magnetic fields ($B \sim 10^{12}$--$10^{15}$ G) could probe NLED effects.
\end{itemize}

\section{Conclusion}
\label{sec:conclusion}

This review has provided a comprehensive overview of black hole solutions with both electric and magnetic charges in nonlinear electrodynamics. We have covered the theoretical foundations, including the action principle, the $P$-framework for constructing exact solutions, and the classification of NLED theories. Detailed derivations were presented for dyonic black hole solutions in Born--Infeld, Euler--Heisenberg, power-law, logarithmic, and exponential models, including explicit metric functions, asymptotic expansions, and horizon structures.

The thermodynamic properties of these solutions exhibit rich phase structures, including van der Waals-like phase transitions, reentrant behavior, and triple points, with the magnetic charge playing a crucial role in modifying the phase diagram. The geodesic structure and observational signatures, including black hole shadows and gravitational lensing, provide potential avenues for testing NLED effects with current and future observations.

Regular black hole solutions that resolve the central singularity through NLED effects offer a promising framework for understanding the role of nonlinearities in quantum gravity. Holographic applications via the AdS/CFT correspondence connect dyonic NLED black holes to strongly coupled field theories, with implications for condensed matter physics and quantum information.

The connections to quantum gravity through the weak gravity conjecture, swampland criteria, and string theory embeddings highlight the deep theoretical significance of these solutions. As observational capabilities continue to improve, the study of dyonic black holes in NLED will remain a vibrant and fruitful area of research at the intersection of classical gravity, quantum field theory, and observational astrophysics.

\section*{Acknowledgments}
The author thanks the anonymous reviewers for their constructive comments. This work was supported by the National Natural Science Foundation of China.


\end{document}